\documentclass[prb]{revtex4}% Physical Review B

\usepackage{graphicx}% Include figure files
\usepackage{dcolumn}% Align table columns on decimal point
\usepackage{bm}% bold math

\begin{document}
\draft
\title{Many-body Green's function theory of ferromagnetic Heisenberg systems with
single-ion anisotropies in more than one direction}
\author{Huai-Yu Wang$^a$}
\address{Department of Physics, Tsinghua University, Beijing 100084, China}
\author{Zhen-Hong Dai}
\address{CCAST(World Lab.), P.\ O. Box 8730, Beijing 100080 and\\
Department of Physics, Tsinghua University, Beijing 100084, China}
\author{P. Fr\"{o}brich$^{b,c}$, P. J. Jensen$^c$ and P. J. Kuntz$^b$}
\address{$^b$Hahn-Meitner-Institut Berlin, Glienicker Stra$\beta $e 100, 14109\\
Berlin, Germany\\
$^c$Institut f\"{u}r Theoretische Physik, Freie Universit\"{a}t Berlin,\\
Arnimallee 14, 14195 Berlin, Germany}
\date{\today }
%\maketitle

\begin{abstract}
The behaviour of ferromagnetic systems with single-ion anisotropies and
magnetic fields in more than one direction is investigated with many-body
Green's function theory generalizing earlier work with uniaxial anisotropies
only. It turns out to be of advantage to construct Green's functions in
terms of the spin operators $S^x,S^y$ and $S^z$, instead of the commonly
used $S^{+},S^{-}$ and $S^z$ operators. The exchange energy terms are
decoupled by RPA and the single-ion anisotropy terms by a generalization of
the Anderson-Callen decoupling. We stress that in the derivation of the
formalism none of the three spatial axes is special, so that one is always
able to select a reference direction along which a magnetization component
is not zero. Analytical expressions are obtained for all three components of
the magnetization and the expectation values $\langle (S^x)^2\rangle $, $%
\langle (S^y)^2\rangle $ and $\langle (S^z)^2\rangle $ for any spin quantum
number $S$. The formalism considers both in-plane and out-of-plane
anisotropies. Numerical calculations illustrate the behaviour of the
magnetization for 3-dimensional and 2-dimensional systems for various
parameters. In the 2-dimensional (monolayer) case, the magnetic
dipole-dipole coupling is included, and a comparison is made between
in-plane and out-of-plane anisotropies.
\end{abstract}


\narrowtext
\maketitle
\section{Introduction}

The spontaneous magnetization of bulk ferromagnets has been successfully
described by a Heisenberg model within the framework of many-body Green's
function theory \cite{tyab,tahir1,callen}, where the Tyablikov decoupling
(random phase approximation: RPA) is the crucial approximation for the
exchange energy term. Usually, only the $z$-component of the magnetization $%
\langle S^z\rangle $ was considered. Callen\cite{callen} developed a
formalism which lead to a closed-form expression for the magnetization $%
\langle S^z\rangle $ in ferromagnetic (FM) systems for any spin quantum
number $S$. This theory has been helpful in studying various systems
described by Heisenberg Hamiltonians \cite{why2,why3,why4,guo}.

Recently, experimental work has provided an incentive to calculate more than
one component of the magnetization. For example, under the influence of an
external field perpendicular to the anisotropy axis, the magnetization
rotates with the variation of the field strength. When a FM layer is covered
by an anti-ferromagnetic layer, the magnetizations across the interface may
have different orientations \cite{moran}. Moreover, the antiferromagnetic
sublattice magnetization may be non-collinear \cite{Ko97}. In FM ultrathin
films, the magnetization may rotate when the temperature or film thickness
are varied; see e.g.\cite{farle,usadel,fro1,fro2}.

Fr\"{o}brich et al. \cite{fro1,fro2} have developed a method using Green's
function theory to calculate more than one component of the magnetization
for FM films. They investigated the reorientation of the magnetization of FM
films caused by an external field and the dipole-dipole coupling by applying
RPA to the exchange energy term and the Anderson-Callen decoupling \cite{ac}
to an {\em out-of-plane} single-ion anisotropy. Later an exact treatment of
the single-ion anisotropy term was proposed by going to higher-order Green's
functions \cite{fro3}. This exact treatment and a comparison with Quantum
Monte Carlo calculations \cite{hen02} showed that the Anderson-Callen
decoupling is a good approximation when the single-ion anisotropy is small
compared to the exchange interaction. The Anderson-Callen decoupling can
therefore often be a useful approximation in this situation, particularly as
it is much simpler to apply than the exact treatment. In Ref. \cite{why1},
general analytical formulas were derived on the basis of RPA and the
Anderson-Callen decoupling for the three components of the magnetization for
3D and 2D FM systems for any spin quantum number. The Green's function
formalism was also applied to uniaxial out-of-plane and uniaxial in-plane
{\em exchange} anisotropies \cite{fro4,Je03,fro5}.

The papers above [12-17] dealt only with a uniaxial {\em out-of-plane}
single-ion anisotropy. In the present paper, we generalize the model by
introducing single-ion anisotropies in all directions of space, thus
obtaining the in-plane and out-of-plane anisotropies as special cases, and
also allowing for a reorientation of the magnetization from out-of-plane to
in-plane and vice versa, depending on the chosen parameters. The new key
features are the formulation in terms of Green's functions defined in the
spin operators $S^x,S^y,S^z$ instead of the $S^{+},S^{-},S^z$ operators, and
the generalization of the Anderson-Callen decoupling to the terms coming
from the single-ion anisotropies for all directions in space. The new
formulation is advantageous if several non-vanishing components of the
magnetization are present.

The paper is organized as follows. Section 2 presents the general formalism
for any spin quantum number. Section 3 shows numerical results for the 3D
case and for a monolayer film (2D). For the latter case we also discuss the
effects of the magnetic dipole-dipole coupling. Section 4 contains our
conclusions.

\section{The formalism}

The Hamiltonian for a ferromagnetic system is
\begin{equation}
H=-\frac J2\sum_{(i,j)}{\bf S}_i\cdot {\bf S}_j-%
\sum_i[K_{2x}(S_i^x)^2+K_{2y}(S_i^y)^2+K_{2z}(S_i^z)^2]-\sum_i{\bf B}\cdot
{\bf S}_i\ \text{.}  \label{1}
\end{equation}
The first term is the Heisenberg exchange interaction ($J>0$). The second
term describes single-ion anisotropies in all three directions of space
where we assume that the anisotropy constants are greater or equal to zero, $%
K_{2\alpha}\geq 0$ for $\alpha=x,y,z$. Note, that due to the exact relation $%
(S^x)^2+(S^y)^2+(S^z)^2=S(S+1)$ only two of the $K_{2\alpha}$ are
independent. The third term is the Zeeman energy, arising from an external
magnetic field.

We use retarded many-body Green's functions defined in accordance with
Bogolyubov and Tyablikov \cite{bog} as
\begin{equation}
G_{ij,\eta }(t-t^{\prime })=\langle \langle A_i;B_j\rangle \rangle _\eta
=-i\theta (t-t^{\prime })\langle [A_i,B_j]_\eta \rangle ,  \label{2}
\end{equation}
where the operators $A_i$ and $B_j$ depend on the problem under
consideration, and the subscripts $i$ and $j$ label lattice sites. $\theta
(t-t^{\prime })$ is the Heavyside function, and $\langle ...\rangle $
denotes the thermodynamical expectation value. Furthermore

\begin{equation}
\lbrack A_i,B_j]_\eta =A_iB_j+\eta \text{ }B_jA_i=F_{ij,\eta} \text{, }\ \ \
\eta =\pm 1.  \label{3}
\end{equation}
After a Fourier transformation to energy space, we obtain the equation of
motion:

\begin{equation}
\omega \langle \langle A_i;B_j\rangle \rangle_{\eta,\omega} =\langle
[A_i,B_j]_\eta \rangle +\langle \langle [A_i,H];B_j\rangle
\rangle_{\eta,\omega} .  \label{4}
\end{equation}

The Green's function is further Fourier transformed to momentum space
\begin{equation}
g({\bf k},\omega )=\frac 1N\sum_{ij}G_{ij}(\omega )e^{i{\bf k\cdot (i-j)}}.
\label{5}
\end{equation}
The aim is to obtain the Green's function $g({\bf k},\omega )$ from a
solution of the equations of motion. Then, statistical averages of the
product of the two operators $B_j$ and $A_i$ (correlation functions), may be
cast in terms of correlations in momentum space, $C_{{\bf k}}$:
\begin{equation}
\langle B_jA_i\rangle =\frac 1N\sum_{{\bf k}}e^{-i{\bf k\cdot (i-j)}}C_{{\bf %
k}}.  \label{6}
\end{equation}
The $C_{{\bf k}}$ can be calculated by the well-known spectral theorem\cite
{tyab}
\begin{equation}
C_{{\bf k}}=\frac i{2\pi }\int \frac{d\omega }{e^{\omega /k_BT}+\eta }[g(%
{\bf k},\omega +i0^{+})-g({\bf k},\omega -i0^{+})],  \label{7}
\end{equation}
where $T$ is the temperature and $k_B$ is the Boltzmann constant.

In order to solve the equations of motion, the higher-order Green's
functions appearing in Eq. (4) have to be decoupled. The following
derivations make use of Fermionic Green's functions ($\eta =+1$) instead of
the Bosonic Green's functions ($\eta =-1$) of earlier work, where zero
eigenvalues of the equation of motion matrix have to be treated carefully,
e.g. \cite{fro1,fro6}.

In previous work [12-19], the operators $A$ and $B$ were the spin operators:
\begin{equation}
A=S^\kappa ,\kappa =+,-,z\text{, and }B=(S^{-})^m(S^z)^n,  \label{8}
\end{equation}
where $m$ and $n$ are zero or positive integers, necessary for treating
larger spin values $S$. The $z$-axis was chosen to be the particular
direction in which the single-ion anisotropy is active.

In this paper, in order to to treat the three directions in space on an
equal footing, we use the following spin operators to construct the Green's
functions:
\begin{equation}
A=S^\alpha ,\alpha =x,y,z\text{, and }B=(S^x)^l(S^y)^m(S^z)^n\text{,}
\label{9}
\end{equation}
where $l,m,n$ are integers or zero.

The advantage of such operators is that the $S^\alpha $'s are equivalent to
each other. A particular direction in space is not preferred.

Analogous to the usual formulation, we apply RPA to the terms coming from
the exchange interaction in the new variables. This corresponds to the
following factorization:
\begin{equation}
\langle \langle S_i^\alpha S_l^\beta ;B_j\rangle \rangle =\langle S_i^\alpha
\rangle \langle \langle S_l^\beta ;B_j\rangle \rangle +\langle S_l^\beta
\rangle \langle \langle S_i^\alpha ;B_j\rangle \rangle \text{, }\alpha \text{%
, }\beta =x,y,z\text{, and }i\neq l.  \label{10}
\end{equation}
RPA is not an appropriate decoupling for the on-site terms coming from the
single-ion anisotropy. For these, we use a generalization of the
Anderson-Callen decoupling (originally derived only for the uniaxial
out-of-plane anisotropies \cite{ac}) by assuming that the decoupling form
factor $\Phi _\alpha $ has an identical form in all spatial directions ($%
\alpha =x,y,z$).

\begin{equation}
\langle \langle S_i^\alpha S_i^\beta +S_i^\beta S_i^\alpha ;B_j\rangle
\rangle =\langle S_i^\alpha \rangle \Phi _\alpha \langle \langle S_i^\beta
;B_j\rangle \rangle =\langle S_i^\beta \rangle \Phi _\beta \langle \langle
S_i^\alpha ;B_j\rangle \rangle \text{, }\alpha \text{, }\beta =x,y,z\text{
and }\alpha \neq \beta \text{.}  \label{11}
\end{equation}
Here
\begin{equation}
\Phi _\alpha =2\left( 1-\frac 1{2S^2}[S(S+1)-\langle S_i^\alpha S_i^\alpha
\rangle ]\right) \text{, }\alpha =x,y,z\text{.}  \label{12}
\end{equation}

As appropriate anti-commutator Green's functions we choose
\begin{equation}
{\bf G}=\left(
\begin{tabular}{l}
$\langle \langle S_i^x;B_j\rangle \rangle _{+1}$ \\
$\langle \langle S_i^y;B_j\rangle \rangle _{+1}$ \\
$\langle \langle S_i^z;B_j\rangle \rangle _{+1}$%
\end{tabular}
\right) \text{.}  \label{13}
\end{equation}
Applying the decoupling procedures (10, 11), we obtain an equation of motion
from which the Green's functions can be determined
\begin{equation}
(\omega {\bf I}-{\bf P}){\bf g}={\bf F}_{+1}^{ij}=\left(
\begin{tabular}{l}
$\langle [S_i^x;B_j]_{+1}\rangle $ \\
$\langle [S_i^y;B_j]_{+1}\rangle $ \\
$\langle [S_i^z;B_j]_{+1}\rangle $%
\end{tabular}
\right) \text{.}  \label{14}
\end{equation}
One of the main advantages of the present formulation is that the matrix $%
{\bf P}$ turns out to be Hermitian, whereas in earlier work \cite
{fro1,fro2,fro3,why1}, the corresponding matrix at this stage of the
decoupling was real and non-Hermitian.
\begin{equation}
{\bf P}=\left(
\begin{tabular}{lll}
$0$ & $iH_z$ & $-iH_y$ \\
$-iH_z$ & $0$ & $iH_x$ \\
$iH_y$ & $-iH_x$ & $0$%
\end{tabular}
\right) \text{. }  \label{15}
\end{equation}
Here
\begin{equation}
H_\alpha =\langle S^\alpha \rangle (J_0-J_{{\bf k}})+B_\alpha +K_{2\alpha
}\langle S^\alpha \rangle \Phi _\alpha \text{, }\ \ \alpha =x,y,z\text{.}
\label{16}
\end{equation}
The quantities $J$'s are defined as
\begin{equation}
J_{{\bf k}}=J\sum_je^{i{\bf k\cdot a}_j}\text{ and }J_0=\gamma J,  \label{17}
\end{equation}
where the summation is over the nearest neighbours and $\gamma $ is the
number of nearest neighbours.

Because the matrix {\bf P} is Hermitian, it has real eigenvalues:
\begin{equation}
\omega _1=0\text{, }\omega _{2,3}=\pm \sqrt{H_x^2+H_y^2+H_z^2}= \pm E_{{\bf k%
}}\text{.}  \label{18}
\end{equation}
The eigenvector matrix can be determined analytically, where the columns
correspond to the eigenvalues $0,-E_{{\bf k}},+E_{{\bf k}}$, respectively
\begin{equation}
{\bf U}=\left(
\begin{array}{ccc}
h_x/h_z & -(h_z-ih_xh_y)/(h_x+ih_yh_z) & -(h_z+ih_xh_y)/(h_x-ih_yh_z) \\
h_y/h_z & -i(h_x^2+h_z^2)/(h_x+ih_yh_z) & i(h_x^2+h_z^2)/(h_x-ih_yh_z) \\
1 & 1 & 1
\end{array}
\right) \text{,}  \label{19}
\end{equation}
and its inverse is
\begin{equation}
{\bf U}^{-1}=\frac 12\left(
\begin{array}{ccc}
2h_xh_z & 2h_yh_z & 2h_z^2 \\
-h_xh_z-ih_y & -h_yh_z+ih_x & h_x^2+h_y^2 \\
-h_xh_z+ih_y & -h_yh_z-ih_x & h_x^2+h_y^2
\end{array}
\right) \text{,}  \label{20}
\end{equation}
where we have defined
\begin{equation}
h_\alpha =\frac{H_\alpha }{E_{{\bf k}}}\text{, }\alpha =x,y,z\text{.}
\label{21}
\end{equation}
The Green's function is obtained by \cite{why1}
\begin{equation}
{\bf g}_{\mu \nu }=\stackrel{3}{\sum_{\lambda =1}}\stackrel{3}{\sum_{\tau =1}%
}\frac{U_{\mu \tau }U_{\tau \lambda }^{-1}}{\omega -\omega _\tau }%
F_{+1}^{\lambda \nu }=\stackrel{3}{\sum_{\lambda =1}}{\bf R}_{\mu \lambda,
{\bf k}}F_{+1}^{\lambda \nu }\ .  \label{22}
\end{equation}

Applying the spectral theorem (7), we get the correlation functions (with $%
\eta =+1$) in momentum space
\begin{equation}
{\bf C}_{{\bf k}}=\left(
\begin{array}{c}
\langle BS^x\rangle \\
\langle BS^y\rangle \\
\langle BS^z\rangle
\end{array}
\right) ={\bf R_kF}_{+1,{\bf k}}\text{.}  \label{23}
\end{equation}
The index ${\bf k}$ is introduced to indicate that the inhomogeneity for the
anti-commutator is momentum-dependent, whereas that for the commutator is
not.

Now we exploit the relation between the anti-commutator and commutator
inhomogeneities,
\begin{equation}
{\bf F}_{+1,{\bf k}}={\bf F}_{-1}+2{\bf C}_{{\bf k}},  \label{24}
\end{equation}
to obtain
\begin{equation}
({\bf I}-2{\bf R}_{{\bf k}}){\bf C}_{{\bf k}}={\bf R}_{{\bf k}}{\bf F}_{-1}%
\text{.}  \label{25}
\end{equation}
This equation is written explicitly as follows
\begin{eqnarray}
&&2\left(
\begin{array}{ccc}
0 & ih_z & -ih_y \\
-ih_z & 0 & ih_x \\
ih_y & -ih_x & 0
\end{array}
\right) \left(
\begin{array}{c}
C_{1{\bf k}} \\
C_{2{\bf k}} \\
C_{3{\bf k}}
\end{array}
\right) =  \nonumber \\
&&\left(
\begin{array}{ccc}
\coth (E_{{\bf k}}/2k_BT) & -ih_z & ih_y \\
ih_z & \coth (E_{{\bf k}}/2k_BT) & -ih_x \\
-ih_y & ih_x & \coth (E_{{\bf k}}/2k_BT)
\end{array}
\right) \left(
\begin{array}{c}
F_{-1,1} \\
F_{-1,2} \\
F_{-1,3}
\end{array}
\right) \text{.}  \label{26}
\end{eqnarray}
It should be pointed out that these equations are not independent of each
other: On the left hand side of equations (26), the first row multiplied by $%
h_x$ plus the second row multiplied by $h_y$ plus the third row multiplied
by $h_z$ is equal to zero. Applying the same condition to the right hand
side, we find
\begin{equation}
h_xF_{-1,1}+h_yF_{-1,2}+h_zF_{-1,3}=0\text{.}  \label{27}
\end{equation}
Taking $B$ of Eq. (9) as $=S^x$, $S^y$ and $S^z$, respectively, we get
\begin{equation}
h_y\langle S^z\rangle =h_z\langle S^y\rangle \text{, }h_x\langle S^z\rangle
=h_z\langle S^x\rangle \text{ and }h_x\langle S^y\rangle =h_y\langle
S^x\rangle \text{.}  \label{28}
\end{equation}
Equations (27) and (28) are called regularity conditions because they are
equivalent to similar equations in Ref. \cite{fro1}, which were derived
there from the condition that the commutator Green's functions are regular
at the origin, $\omega =0$.

For the convenience of the discussion below, we now define two quantities $%
Q_\alpha $ and $h_{\alpha \beta }$ as
\begin{equation}
Q_\alpha =\frac 1N\sum_{{\bf k}}\frac{\coth (E_{{\bf k}}/2k_BT)}{h_\alpha }%
\text{ and }h_{\alpha \beta }=\frac{h_\alpha }{h_\beta }=\frac{H_\alpha }{%
H_\beta }\text{, }\ \ \ \alpha \text{, }\beta =x,y,z\text{.}  \label{29}
\end{equation}
Before further discussion, we recall that Callen \cite{callen} derived a
general expression for calculating $\langle S^z\rangle $ only:
\begin{equation}
\langle S^z\rangle =\frac{(S-Q)(Q+1)^{2S+1}+(S+1+Q)Q^{2S+1}}{%
(Q+1)^{2S+1}-Q^{2S+1}}\text{,}  \label{30}
\end{equation}
where $S$ is the spin quantum number and $Q=\frac 1N\sum_{{\bf k}%
}1/(e^{\omega ({\bf k)}/k_BT}-1)$ with $\omega ({\bf k)}$ being the
dispersion relation. We found \cite{why1} that putting Eq. (30) into a
symmetric form has some formal advantage
\begin{equation}
\langle S^z\rangle =\frac{(2S+1-Q_1)(Q_1+1)^{2S+1}+(2S+1+Q_1)(Q_1-1)^{2S+1}}{%
2[(Q_1+1)^{2S+1}-(Q_1-1)^{2S+1}]}\text{,}  \label{31}
\end{equation}

where $Q_1=\frac 1N\sum_{{\bf k}}\coth(\omega ({\bf k)}/2k_BT)=1+2Q$.

Now we are prepared to derive the required formulas for the components of
the magnetization from Eqs. (26) and (27). We deal first with the case of an
in-plane magnetization, say in the $xy$-plane. It is comparatively easy to
get general expressions for this case. Then we proceed to the most general
case, where it does not matter in which direction the magnetization points.

\subsection*{2.1. Magnetization in the $xy$-plane.}

Here we take the magnetization to be in the $xy$-plane. In this case $%
\langle S^z\rangle =0$ and $h_z=0$, and we assume that $h_x\neq 0$, i.e. we
select the $x$-direction as reference direction. Two of the Eqs. (26) can be
rewritten as
\begin{eqnarray}
-2C_3 &=&iQ_xF_{-1,2}+F_{-1,3}\text{,}  \nonumber \\
-2h_{yx}C_1+2C_2 &=&h_{yx}F_{-1,1}-F_{-1,2}+iQ_xF_{-1,3}\text{.}  \label{32}
\end{eqnarray}
These equations allow us to obtain the expressions for the magnetization and
all correlations for each spin value $S$ by specifying the proper $%
B=(S^x)^l(S^y)^m(S^z)^n$ operators, and using the regularity conditions and
some spin algebra. Because of the Anderson-Callen decoupling (11, 12), the
correlations $\langle (S^x)^2\rangle $, $\langle (S^y)^2\rangle $ and $%
\langle (S^z)^2\rangle $ are also needed. We list in Table I the explicit
expressions for $\langle S^x\rangle $, $\langle (S^x)^2\rangle $ and $%
\langle (S^y)^2\rangle $ for spin quantum numbers $S=1/2$ to 5/2, where
\begin{equation}
P^2=1+h_{yx}^2\text{.}  \label{33}
\end{equation}
Note that this quantity is momentum independent due to the regularity
conditions (28), $h_{yx}=\langle {S^y}\rangle /\langle {S^x}\rangle $.

Table I. The analytical expressions $\langle S^x\rangle $, $\langle
(S^x)^2\rangle $ and $\langle (S^y)^2\rangle $ and for $S=1/2$ to 5/2.
\[
\begin{array}{cccc}
S & \langle S^x\rangle & \langle (S^x)^2\rangle & \langle (S^y)^2\rangle \\
1/2 & 1/2Q_x & 1/4 & 1/4 \\
1 & \frac{4Q_x}{3Q_x^2+P^2} & \frac{2(Q_x^2+1)}{3Q_x^2+P^2} & \frac{%
2(Q_x^2+h_{yx}^2)}{3Q_x^2+P^2} \\
3/2 & \frac{5Q_x^2+P^2}{2Q_x(Q_x^2+P^2)} & \frac{5Q_x^2+12+P^2}{4(Q_x^2+P^2)}
& \frac{5Q_x^2+R^2+12h_{yx}^2}{4(Q_x^2+P^2)} \\
2 & \frac{4Q_x(5Q_x^2+3P^2)}{5Q_x^4+10Q_x^2P^2+P^4} & \frac{%
10Q_x^4+42Q_x^2+6P^2+6Q_x^2P^2}{5Q_x^4+10Q_x^2P^2+P^4} & \frac{%
10Q_x^4+42Q_x^2h_{yx}^2+6P^2h_{yx}^2+6Q_x^2P^2}{5Q_x^4+10Q_x^2P^2+P^4} \\
5/2 & \frac{35Q_x^4+42Q_x^2P^2+3P^4}{2Q_x(3Q_x^4+10Q_x^2P^2+3R^4)} & \frac{%
35Q_x^4+14Q_x^2(3P^2+16)+3(P^2+32)P^2}{4(3Q_x^4+10Q_x^2P^2+3P^4)} & \frac{%
35Q_x^4+14Q_x^2(3P^2+16h_{yx}^2)+3(P^2+32h_{yx}^2)P^2}{%
4(3Q_x^4+10Q_x^2P^2+3P^4)}
\end{array}
\]

For $P^2=1$ in the expression for $\langle S^x\rangle $ in Table I one
obtains the same result as from Eq. (31) when substituting $S=1/2$ to 5/2
into it and replacing $z$ by $x$.

With $P^2=1+h_{yx}^2$, a general expression analogous to Eq. (31) is found

\begin{equation}
\langle S^x\rangle =\frac{%
[(2S+1)P-Q_x](Q_x+P)^{2S+1}+[(2S+1)P+Q_x](Q_x-P)^{2S+1}}{%
2P^2[(Q_x+P)^{2S+1}-(Q_x-P)^{2S+1}]}.  \label{34}
\end{equation}
Substituting $S=1/2$ to 5/2 into Eq.(34), we obtain the expressions in the
column 2 in Table I. Thus we suggest that Eq.(34) is the required correct
expression for any $S$ (we have no analytical proof for the general formula).

We also suggest the following analytical expressions for the three
correlations for any spin value
\begin{equation}
\langle (S^x)^2\rangle =\frac{2S(S+1)+Q_xS^x(P^2-3)}{2P^2}\text{,}
\label{35}
\end{equation}
\begin{equation}
\langle (S^y)^2\rangle =\frac{2S(S+1)h_{yx}^2+(P^2-3h_{yx}^2)Q_xS^x}{2P^2}%
\text{,}  \label{36}
\end{equation}
and
\begin{equation}
\langle (S^z)^2\rangle =\frac 12Q_xS^x\text{.}  \label{37}
\end{equation}
It is seen that the exact relation
\begin{equation}
\langle (S^x)^2\rangle +\langle (S^y)^2\rangle +\langle (S^z)^2\rangle
=S(S+1)  \label{38}
\end{equation}
is satisfied. Also, when $S=1/2$ to 5/2 is substituted into Eqs.(35, 36), we
obtain the expressions in columns 3 and 4 in Table I.

Because in the present case the $x$ and $y$-directions should be equivalent,
exchanging $x\longleftrightarrow y$ in Eqs.(34-37) and in Table I must also
yield valid expressions.

\subsection*{2.2. The general case}

We now take the $z$-direction as reference direction, supposing $h_z\neq 0$,
and taking two equations from Eqs.(26)
\begin{equation}
\begin{tabular}{l}
$2C_2-2h_{yz}C_3=-iQ_zF_{-1,1}-F_{-1,2}+h_{yz}F_{-1,3}\text{,}$ \\
$-2C_1+h_{xz}C_3=F_{-1,1}-iQ_zF_{-1,2}-h_{xz}F_{-1,3}\text{.}$%
\end{tabular}
\label{39}
\end{equation}
Again, by specifying proper B operators, one obtains from these equations
analytical expressions for the magnetization components and all correlations
for each $S$. In Table II the explicit expressions for $\langle S^z\rangle $%
, $\langle (S^z)^2\rangle $, $\langle (S^x)^2\rangle $ and $\langle
(S^y)^2\rangle $ for spin quantum number $S=1/2$ to 5/2 are listed, where
\begin{equation}
R^2=1+h_{xz}^2+h_{yz}^2\text{.}  \label{40}
\end{equation}
Table II. The analytical expressions for $\langle S^z\rangle $, $\langle
(S^z)^2\rangle $, $\langle (S^x)^2\rangle $ and $\langle (S^y)^2\rangle $
for $S=1/2$ to 5/2.
\[
\begin{array}{ccc}
S & \langle S^z\rangle & \langle (S^z)^2\rangle \\
1/2 & 1/2Q_z & 1/4 \\
1 & \frac{4Q_z}{3Q_z^2+R^2} & \frac{2(Q_z^2+1)}{3Q_z^2+R^2} \\
3/2 & \frac{5Q_z^2+R^2}{2Q_z(Q_z^2+R^2)} & \frac{5Q_z^2+R^2+12}{4(Q_z^2+R^2)}
\\
2 & \frac{4Q_z(5Q_z^2+3R^2)}{5Q_z^4+10Q_z^2R^2+R^4} & \frac{%
10Q_z^4+6Q_z^2(R^2+7)+6R^2}{5Q_z^4+10Q_z^2R^2+R^4} \\
5/2 & \frac{35Q_z^4+42Q_z^2R^2+3R^4}{2Q_z(3Q_z^4+10Q_z^2R^2+3R^4)} & \frac{%
35Q_z^4+14Q_z^2(3R^2+16)+3(R^2+32)R^2}{4(3Q_z^4+10Q_z^2R^2+3R^4)}
\end{array}
\]
\[
\begin{array}{ccc}
S & \langle (S^x)^2\rangle & \langle (S^y)^2\rangle \\
1/2 & 1/4 & 1/4 \\
1 & \frac{2(Q_z^2+h_{xz}^2)}{3Q_z^2+R^2} & \frac{2(Q_z^2+h_{yz}^2)}{%
3Q_z^2+R^2} \\
3/2 & \frac{5Q_z^2+R^2+12h_{xz}^2}{4(Q_z^2+R^2)} & \frac{%
5Q_z^2+R^2+12h_{yz}^2}{4(Q_z^2+R^2)} \\
2 & \frac{10Q_z^4+6Q_z^2(R^2+7h_{xz}^2)+6R^2h_{xz}^2}{5Q_z^4+10Q_z^2R^2+R^4}
& \frac{10Q_z^4+6Q_z^2(R^2+7h_{yz}^2)+6h_{yz}^2R^2}{5Q_z^4+10Q_z^2R^2+R^4}
\\
5/2 & \frac{35Q_z^4+14Q_z^2(3R^2+16h_{xz}^2)+3(R^2+32h_{xz}^2)R^2}{%
4(3Q_z^4+10Q_z^2R^2+3R^4)} & \frac{%
35Q_z^4+14Q_z^2(3R^2+16h_{yz}^2)+3(R^2+32h_{yz}^2)R^2}{%
4(3Q_z^4+10Q_z^2R^2+3R^4)}
\end{array}
\]

Similar to the case in subsection 2.1, we suggest that these quantities can
be expressed by the following equations for any $S$.
\begin{equation}
\langle S^z\rangle =\frac{%
[(2S+1)R-Q_z](Q_z+R)^{2S+1}+[(2S+1)R+Q_z](Q_z-R)^{2S+1}}{%
2R^2[(Q_z+R)^{2S+1}-(Q_z-R)^{2S+1}]}\text{,}  \label{41}
\end{equation}
\begin{equation}
\langle (S^z)^2\rangle =\frac{2S(S+1)+(R^2-3)Q_zS^z}{2R^2}\text{,}
\label{42}
\end{equation}
\begin{equation}
\langle (S^x)^2\rangle =\frac{2S(S+1)h_{xz}^2+(R^2-3h_{xz}^2)Q_zS^z}{2R^2}%
\text{, }  \label{43}
\end{equation}
and
\begin{equation}
\langle (S^y)^2\rangle =\frac{2S(S+1)h_{yz}^2+(R^2-3h_{yz}^2)Q_zS^z}{2R^2}%
\text{.}  \label{44}
\end{equation}
Here equation (38) is again satisfied. As long as $\langle S^z\rangle $ is
calculated, the other two components of the magnetization can be computed
from the regularity conditions Eqs.(28).

Because none of the three directions is special, we can also pick the $x$-
or $y$-direction as reference direction. Therefore, in Eqs. (41-44) and in
Table II, the formulas remain valid when performing the exchanges of
coordinates, e.g., $x\rightarrow z\rightarrow y\rightarrow x$. Note too,
that the expressions of Eqs. (41--44) revert to those of Eqs. (34--37) of
subsection 2.1 after making the exchange $x\rightarrow y\rightarrow
z\rightarrow x$ and setting $h_{zx}=0$.

\subsection*{2.3. The effective (temperature-dependent) anisotropies}

The absolute value of the magnetization and the equilibrium angles $\theta
_0 $ and $\varphi _0$ of the magnetization direction are determined by the
following equations:
\begin{equation}
M^2(T)=\langle S^x\rangle ^2+\langle S^y\rangle ^2+\langle S^z\rangle ^2%
\text{,}  \label{45}
\end{equation}
\begin{equation}
\theta _0=\arctan \frac{\sqrt{\langle S^x\rangle ^2+\langle S^y\rangle ^2}}{%
\langle S^z\rangle }\text{, }\varphi _0=\arctan \frac{\langle S^y\rangle }{%
\langle S^x\rangle }\text{. }  \label{46}
\end{equation}
Fr\"{o}brich et al. \cite{fro1} have determined the effective
(temperature-dependent) anisotropy coefficients $K_{2z}(T)$
non-perturbatively for the out-of-plane anisotropy case by minimizing the
free energy with respect to the orientation angles of the magnetization. In
this paper, we are also able to determine the coefficients $K_{2x}(T)$ and $%
K_{2y}(T)$ in the same manner.

The free energy corresponding to the Heisenberg model with anisotropies in
each direction reads
\begin{eqnarray}
F(T)&=&F_0(T)-K_{2x}(T)\sin^2\theta\cos^2\varphi-K_{2y}(T)\sin^2\theta\sin^2
\varphi -K_{ 2z}(T)\cos^2\theta  \nonumber \\
&-&B_xM\sin\theta\cos\varphi-B_yM\sin\theta\sin\varphi-B_zM\cos \theta
\label{47}
\end{eqnarray}
We have mentioned that, due to the relation $(S^x)^2+(S^y)^2+(S^z)^2=S(S+1)$%
, the three anisotropies are not independent of each other. Therefore we are
allowed to transform for instance $K_{2x}\rightarrow K_{2x}-K_{2y}$, $%
K_{2y}\rightarrow 0$, and $K_{2z}\rightarrow K_{2z}-K_{2y}$.

Choosing the $xz$-plane as reorientation plane by taking $B_y=0$, which
leads to $\varphi =0$, the variation of $dF/d\theta=0$ yields
\begin{equation}
K_{2z}(T)-K_{2x}(T)=\frac{B_xM\cos\theta_0-B_zM\sin\theta_0} {%
2\sin\theta_0\cos\theta_0 }.  \label{48}
\end{equation}
The result of the uniaxial out-of plane case \cite{fro1} is obtained by
taking $K_{2x}=0$. The angle $\theta_0$ is the equilibrium angle.

When considering for example the in-plane magnetization with $K_{2x}\neq 0$
and $K_{2y}\neq 0$ and $K_{2z}=0$ (no out-of-plane anisotropy), one has two
equations,
\begin{equation}
dF(T)/d\theta=0,  \label{49}
\end{equation}
and
\begin{equation}
dF(T)/d\varphi=0  \label{50}
\end{equation}
to determine $K_{2x}(T)$ and $K_{2y}(T)$ with the result:

\begin{eqnarray}
K_{2x}(T)&=&\frac{B_zM}{2\cos\theta_0}-\frac{B_xM} {2\sin\theta_0\cos%
\varphi_0},  \nonumber \\
K_{2y}(T)&=&\frac{B_zM}{2\cos\theta_0}-\frac{ B_yM}{2\sin\theta_0\sin%
\varphi_0},  \label{51}
\end{eqnarray}
where the angles $\theta_0, \varphi_0$ are to be considered as equilibrium
angles.

\section{Numerical results}

In the numerical calculations, the strength of the exchange energy is set to
$J=100$. The parameters are scaled as in Refs. \cite{fro1,fro2},
\begin{equation}
J\rightarrow J/S(S+1)\text{, }B\rightarrow B/S\text{, }K_2\rightarrow
K_2/S(S-1/2).  \label{52}
\end{equation}
Scaling is introduced to obtain magnetization curves which are approximately
universal for different spin values $S$. The scaling of $K_2$ guarantees the
property lim$_{T\rightarrow 0}K_2(T)/K_2=1$. For the 3-dimensional (3D)
case, we use a simple cubic lattice and for the 2-dimensional (2D) case, a
square lattice.

\subsection*{3.1. The 3D case}

For the 3D examples, we always put $K_{2z}=0$ and $K_{2x}=1$. For the
uniaxial in-plane case ($K_{2y}=0$) the magnetization points in the $x$%
-direction. In the absence of magnetic fields, the corresponding scaled
Curie point $T_c/T_c(S=1)$, normalized to that of $S=1$, versus the spin
quantum number $S$ is plotted as squares in Fig.1. The scaled $T_c$ slightly
decreases with increasing $S$.
%1
\begin{figure}[htb]
\begin{center}
\protect
\includegraphics*[bb=0 0 550 770,
angle=-90,clip=true,width=10cm]{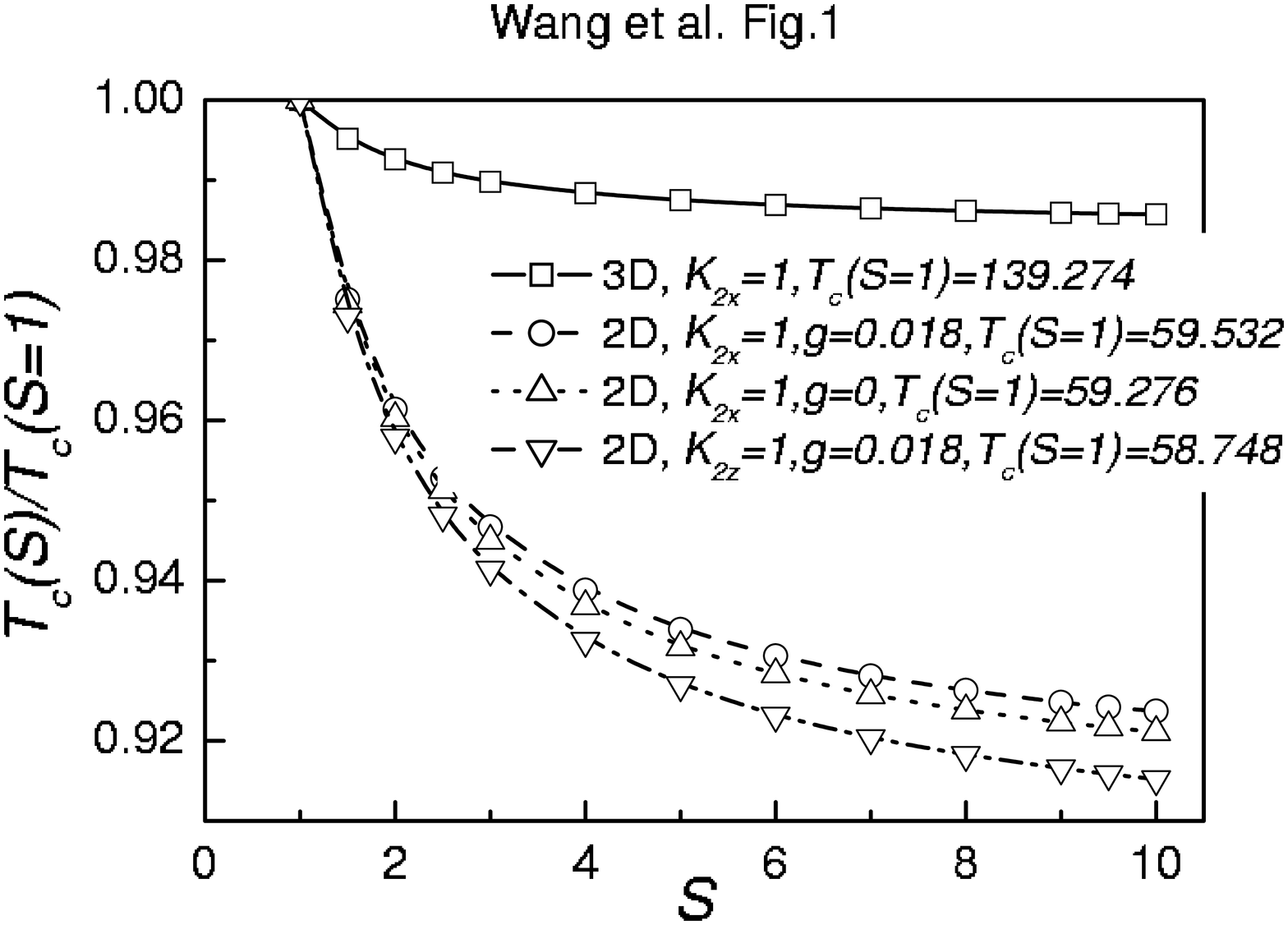}
\protect
\caption{The scaled Curie temperature $T_c$ versus the spin quantum number
$ S $ when there is only a single-ion anisotropy in one direction. Squares: 3D
case. Upward triangles: 2D case without dipole-dipole interaction. Circles
and downward triangles : 2D case with dipole-dipole interaction (with
strength $g=0.018$), when the easy-axis is in plane and out of plane,
respectively. }
\end{center}
\end{figure}
%2
\begin{figure}[htb]
\begin{center}
\protect
\includegraphics*[bb=0 0 550 770,
angle=-90,clip=true,width=12cm]{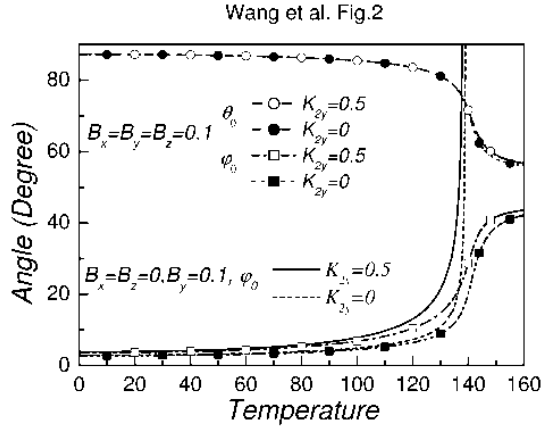}
%\includegraphics*[bb=0 0 55 74,
%angle=-90,clip=true,width=10cm]{Wang2.eps}
\protect
\caption{The angles of the direction of the magnetization ($\theta _0$, $%
\varphi _0)$, defined by Eqs.(\ref{46}), as functions of the temperature for
$K_{2z}=0,K_{2x}=1$ and various $K_{2y}$ anisotropies and magnetic fields.
The symbols are just to label curves.}
\end{center}
\end{figure}

%3
\begin{figure}[htb]
\begin{center}
\protect
\includegraphics*[bb=0 0 550 770,
angle=-90,clip=true,width=10cm]{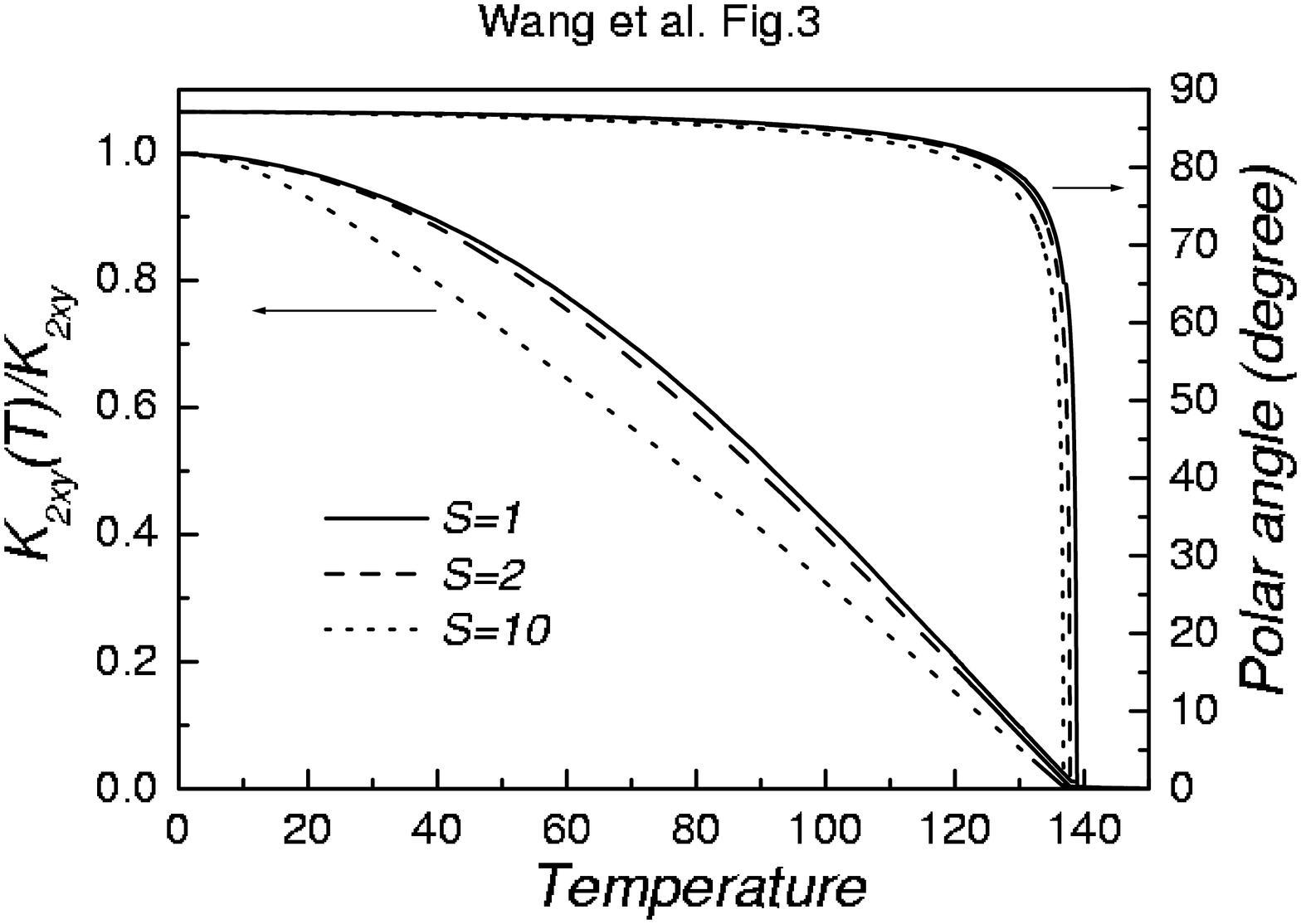}
\protect
\caption{The effective anisotropy $K_{2xy}(T)/K_{2xy}(T=0)$ and the
equilibrium polar angle $\theta _0$ for $S=1$, 2 and 10 in the 3D case for
in-plane anisotropy. The parameters are $K_{2xy}=K_{2x}=K_{2y}=1$ and $%
B_z=0.1$.}
\end{center}
\end{figure}
%4
\begin{figure}[htb]
\begin{center}
\protect
\includegraphics*[bb=0 0 550 770,
angle=0,clip=true,width=12cm]{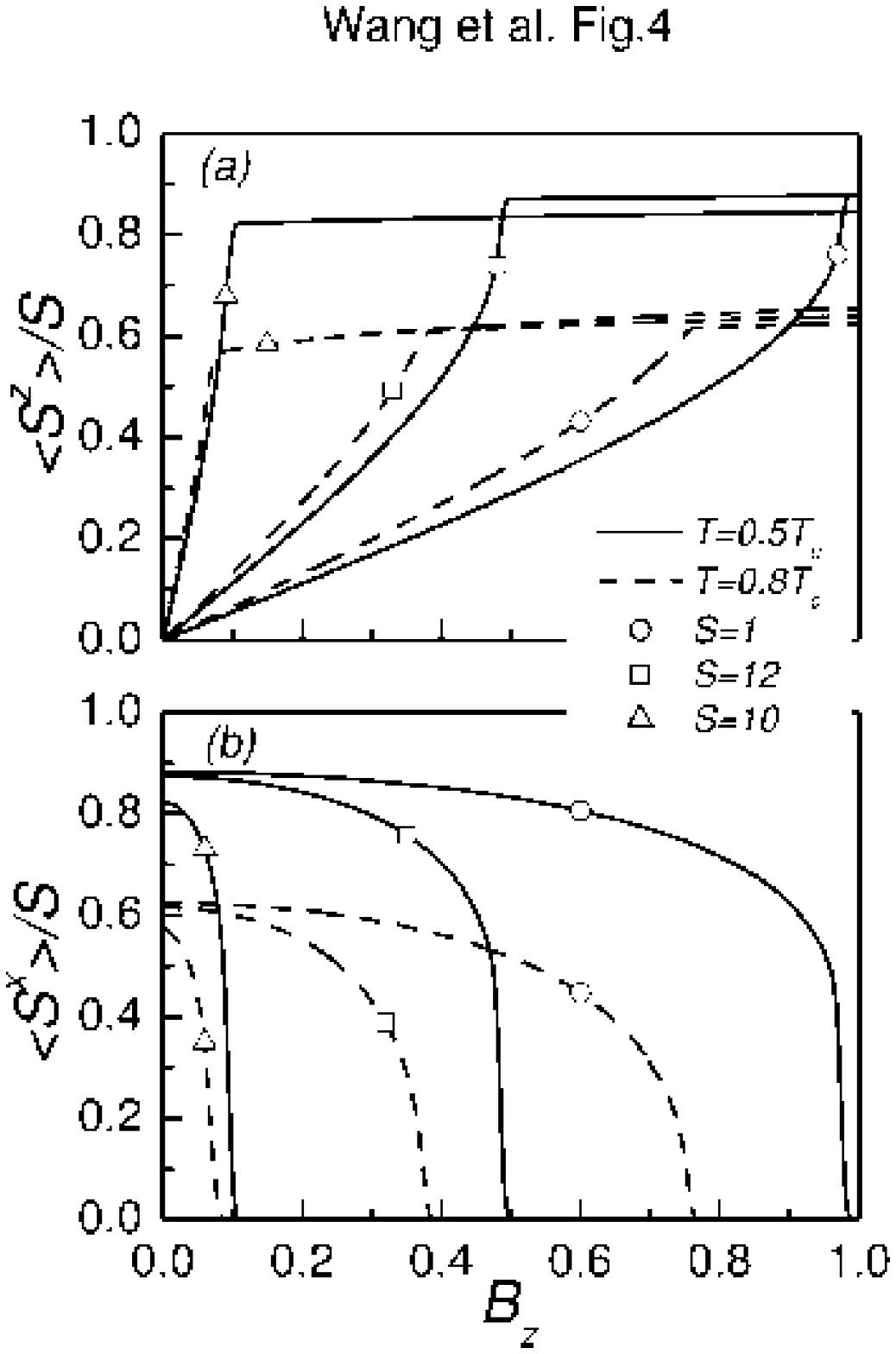}
%\includegraphics*[bb=0 0 60 80,
%angle=0,clip=true,width=10cm]{Wang4.eps}
\protect
\caption{The components of the magnetization (a) $\langle S^z\rangle /S$ and
(b) $\langle S^x\rangle /S$ as functions of the external field in $z$%
-direction $B_z$ for $S=1$, 2 and 10 and at temperatures $T=0.5T_c$ and 0.8$%
T_c$. $T_c$ is the scaled Curie temperature of the 3D case shown by squares
in Fig.1.}
\end{center}
\end{figure}

We show some examples when including a magnetic field ${\bf B}=(B_x,B_y,B_z)$
in addition to the anisotropies $K_{2x}=1,K_{2y}\neq 0,K_{2z}=0$. The
direction of the magnetization will ultimately turn towards the field
direction when the temperature goes above the Curie point (defined from $%
\langle S^x\rangle =0$ for the field-free case). For instance, when ${\bf B}%
=(0.1,0.1,0.1)$, ($\theta _0$, $\varphi _0$) approach (57.74, 45) degrees,
which is the [111] direction, and when ${\bf B}=(0$, 0.1, 0), $\varphi _0$
approaches 90 degrees, as shown in Fig. 2. However, at low temperature,
where the field is not strong enough to overcome the effect of the
anisotropy, the magnetization will be close to the direction along which the
anisotropy is strongest. With increasing temperature, the magnetization
gradually turns towards the field direction. At zero temperature, the
initial angles of $\varphi _0$ are 3.60 and 2.74 degrees for $K_{2y}=0.5$
and 0, respectively, for both applied fields ${\bf B}=(0.1$, 0.1, 0.1) and $%
(0$, 0.1, 0). When $K_{2y}=0.5$, the magnetization turns more easily than
with $K_{2y}=0$. Note that the dashed and dot-dashed lines are very similar.
This reflects the fact that the magnetization component in the plane is
almost the same for the two cases of $K_{2y}=0.5$ and 0. If ${\bf B}=(0.1$,
0.1, 0), the magnetization is in the $xy$-plane ($\theta _0=90^o$). The
curve of the angle $\varphi _0$ versus temperature is identical to the solid
line in Fig. 2 for $K_{2y}=0.5$ and to the dotted line for $K_{2y}=0$.

As a further example we discuss the in-plane anisotropy case, with $%
K_{2xy}=K_{2x}=K_{2y}=1$. The effective anisotropy coefficient as a function
of the temperature can be calculated by Eq. (51). Figure 3 shows examples
for the spin values $S=1$, 2 and 10 calculated for $B_z=0.1$ and $B_x=B_y=0$%
. At zero temperature , $K_{2xy}(T=0)$ is the same as the anisotropy
parameter of the Hamiltonian $K_{2xy}$. With increasing temperature, $%
K_{2xy}(T)$ decreases. Close to the Curie point, the coefficients approach
zero. With the present choice of parameters the reorientation temperature is
close to the Curie point; in general they can be quite different. In the
intermediate temperature range, one observes a sensitivity to the spin
value: the larger the spin $S$, the smaller the effective coefficient. The
larger the spin the more increases the range of a linear behaviour of the
effective anisotropy, which is in accordance with mean field results where
for $S\rightarrow \infty $ or for classical spins one obtains a linear
behaviour $K_2(T)=K_2(0)(1-T/T_C)$ \cite{Millev}. Figure 3 also shows the
corresponding equilibrium angle $\theta _0$ defined by Eq.(46). The in-plane
anisotropy should force the magnetization into the $xy$- plane. However,
because of a small field in $z$-direction $B_z=0.1$, the magnetization
deviates slightly from the plane at zero temperature, and there is a small
initial angle. It is the same for each $S$. With increasing temperature, the
magnetization reorients slowly, owing to the action of the field. Close to
the Curie point, the effective anisotropy approaches zero, so that the
magnetization reorients rapidly in the direction of the $z$-axis, the
direction of the external field.

The orientation angle $\theta _0$ of the magnetization depends on the
temperature as well as on the external field. In Figure 4, we plot the
magnetization components $\langle S^x\rangle /S$ and $\langle S^z\rangle /S$
versus the unscaled external field $B_z$ at two temperatures for $S=1$, 2
and 10. Because of scaling the temperature dependence for different spin
values looks similar. The reorientation fields at which the reorientation to
the field direction has occurred however are different and are given by $%
B_z=0.76$, 0.38 and 0.08, respectively at $T=0.8T_c$. The larger the spin
value $S$, the easier is the reorientation of the magnetization. To compare
the reorientation behaviour at different temperatures, we also show the
results at $T=0.5T_c$. In this case, the magnitudes of the field at which
the magnetization reorientation occurs are $B_z=0.98$, 0.49 and 0.10,
respectively, which are higher than those at temperature $T=0.8T_c$. At a
lower temperature, it is more difficult for the magnetization to turn under
the action of the field, as one expects.

\subsection*{3.2. The 2D-case (monolayer)}

For the discussion of a monolayer (the 2D-case), we add the dipole-dipole
interaction (DI) to the Hamiltonian of Eq.(1)

\begin{eqnarray}
&&\frac g2\sum_{ij}\frac{{\bf S}_i\cdot {\bf S}_j-3({\bf S}_i\cdot {\bf u}%
_{ij})({\bf S}_j\cdot {\bf u}_{ij})}{r_{ij}^3}  \nonumber \\
&=&\sum_{ij}\frac g{2r_{ij}^3}%
\{S_i^xS_j^x[1-3(u_{ij}^x)^2]+S_i^yS_j^y[1-3(u_{ij}^y)^2]+S_i^zS_j^z[1-3(u_{ij}^z)^2]\}%
\text{,}  \label{53}
\end{eqnarray}
where $r_{ij}$ is the distance between sites $i$ and $j$, and its components
along three axes are $r_{ij}u_{ij}^x$, $r_{ij}u_{ij}^y$ and $r_{ij}u_{ij}^z$%
, respectively. The higher-order Green's functions containing DI terms are
also decoupled by RPA, i.e. by Eq.(10). As a result, there appear additive
terms due to DI in $H_\alpha $ of Eq.(16), which include non-dispersive
parts as well as dispersive parts. As an approximation, we neglect the
dispersive parts and retain the non-dispersive parts only. Such an
approximation was used before in Ref.\cite{fro2}. The full RPA treatment of
the dipole coupling would require complex arithmetic numerically (see
Appendix A of Ref.\cite{fro2}). An exception is the uniaxial out-of-plane
case, for which it was shown in this Appendix that the mean-field results
are close to RPA. Using complex arithmetic numerically is not in principle a
hindrance, but is quite tedious to perform. One obtains
\begin{equation}
H_\alpha =\langle S^\alpha \rangle (J_0-J_{{\bf k}})+B_\alpha +K_{2\alpha
}\langle S^\alpha \rangle \Phi _\alpha -g\langle S^\alpha \rangle \sum_l%
\frac 1{r_{0l}^3}[1-3\frac{(u_{0l}^\alpha )^2}{r_{0l}^2}]\text{, }\alpha =x%
\text{, }y\text{, }z\text{.}  \label{54}
\end{equation}
In the last term of Eq.(54), the subscript 0 means the origin, and the
summation covers all lattice sites except for the origin. The lattice sum
can be done and Eq.(54) becomes explicitly
\begin{eqnarray}
H_{x,y} &=&\langle S^{x,y}\rangle (J_0-J_{{\bf k}})+B_{x,y}+K_{2x,y}\langle
S^{x,y}\rangle \Phi _{x,y}+gT_0\langle S^{x,y}\rangle \text{, }  \nonumber \\
H_z &=&\langle S^z\rangle (J_0-J_{{\bf k}})+B_z+K_{2z}\langle S^z\rangle
\Phi _z-2gT_0\langle S^z\rangle \text{,}  \label{55}
\end{eqnarray}
where $T_0=\sum_{l,m}\frac{l^2}{(l^2+m^2)^{5/2}}=4.5165$. To compare with
the results in Ref.\cite{fro2}, we choose $g=0.018$ (a value which
corresponds to Ni with the present set of parameters) in most cases. Note
that the DI strength $g$ is also scaled as

\begin{equation}
g\rightarrow g/S(S+1)\text{.}  \label{56}
\end{equation}
Correspondingly, the expressions for the effective anisotropies Eqs.(48, 51)
will have additional terms coming from the DI, as in Ref. \cite{fro2} for
the out-of-plane case.

Now we proceed to some numerical examples. Owing to the Mermin-Wagner
theorem \cite{MW66}, a Heisenberg monolayer without some kind of anisotropy
has no spontaneous magnetization. A spontaneous magnetization can be
achieved by DI alone. Figure 5 shows the absolute value of the in-plane
magnetization due to the action of DI as function of the temperature. One
observes that for different spin quantum numbers $S$, the scaled Curie
temperature $T_c$ is the same for a fixed DI strength. The inner panel shows
the $T_c-g$ curve in the range of $g=0.01$---1. When $g$ approaches zero, $%
T_c$ decreases drastically, and approaches zero for $g\rightarrow 0$ in
accordance with the Mermin-Wagner theorem.
%5
\begin{figure}[htb]
\begin{center}
\protect
\includegraphics*[bb=0 0 420 350,
angle=0,clip=true,width=10cm]{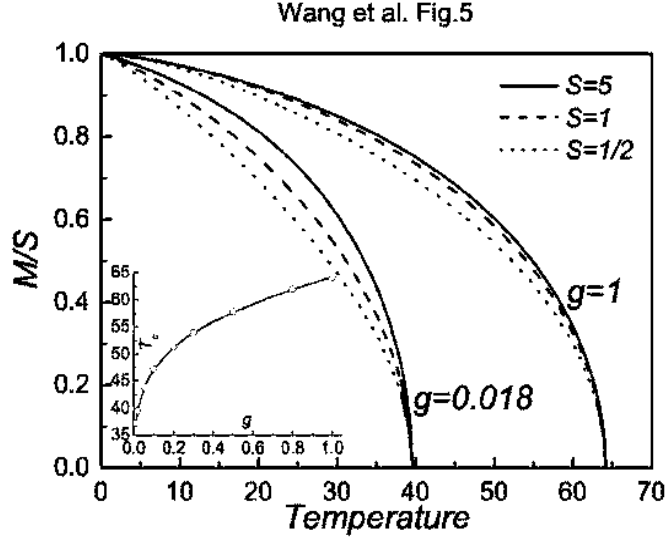}
%\includegraphics*[bb=0 0 299 249,
%angle=0,clip=true,width=10cm]{WangFig5.eps}
\protect
\caption{Magnetization-temperature curve for $S=1$, 2 and 5 for the monolayer
with
DI strength $g=0.018$ and 1, respectively. $M$ is the magnetization defined
in Eq.(\ref{45}) for the in-plane situation. When $g$ is fixed, $T_c$ is same
for any S. The inner panel: $T_c$ as a function of the DI strength $g$.}
\end{center}
\end{figure}
%

%6
\begin{figure}[htb]
\begin{center}
\protect
\includegraphics*[bb=0 0 420 380,
angle=0,clip=true,width=10cm]{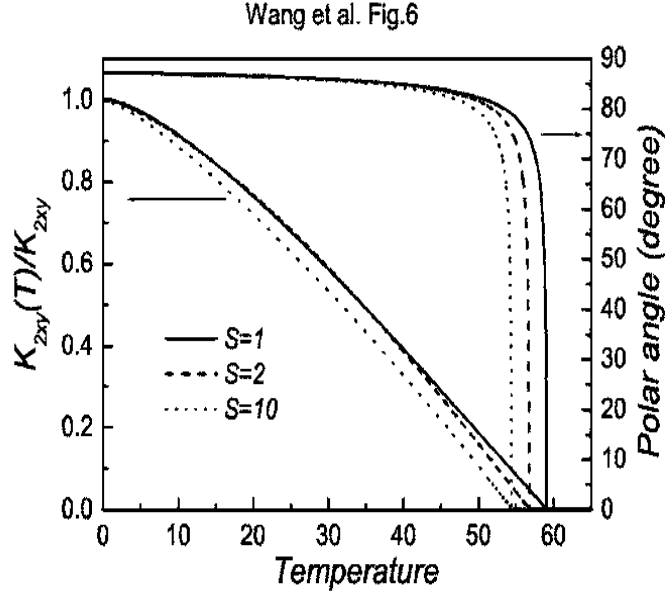}
%\includegraphics*[bb=10 10 334 251,
%angle=0,clip=true,width=10cm]{WangFig6.eps}
\protect
\caption{Effective anisotropy $K_{2xy}(T)/K_{2xy}$ and equilibrium polar
angle $\theta _0$ for $S=1$, 2 and 10 in the 2D case. The parameters are same
as in Fig. 3.}
\end{center}
\end{figure}
%
%7
\begin{figure}[htb]
\begin{center}
\protect
\includegraphics*[bb=0 0 224 267,
angle=0,clip=true,width=10cm]{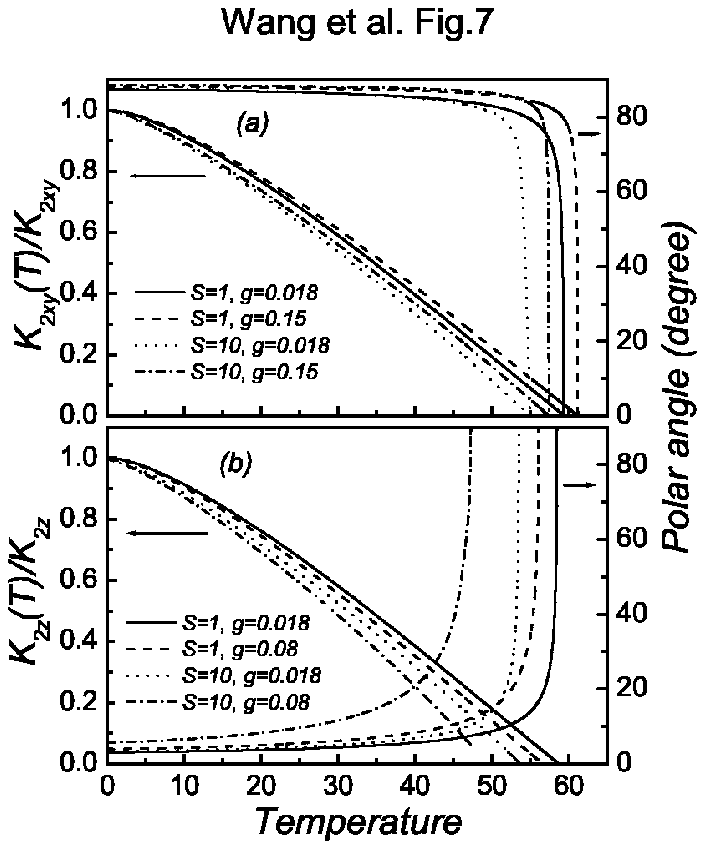}
\protect
\caption{(a) Effective anisotropy $K_{2xy}(T)/K_{2xy}$ and equilibrium polar
angle $\theta _0$ for $S=1$ and 10 for the monolayer with DI strengths $%
g=0.018$ and 0.15 for the in-plane case. The parameters are $K_{2xy}=1$ and $%
B_z=0.1$. (b) Effective anisotropy $K_{2z}(T)/K_{2z}$ and equilibrium polar
angle $\theta _0$ for $S=1$ and 10 for the monolayer with DI strengths $%
g=0.018$ and 0.08 for the out-of-plane case. The parameters are $K_{2z}=1$
and $B_x=0.1$.}
\end{center}
\end{figure}
%8
\begin{figure}[htb]
\begin{center}
\protect
\includegraphics*[bb=0 0 51 72,
angle=-90,clip=true,width=10cm]{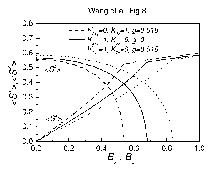}
\protect
\caption{The components of the magnetization $\langle S^x\rangle $ and $%
\langle S^z\rangle $ as functions of the external field for $S=1$ at the
temperature $T/T_c=0.8$. The dashed lines correspond to the out-of-plane
anisotropy with a field $B_x$ in $x$-direction. The solid and dotted lines
correspond to the in-plane anisotropy with a field $B_z$ in z-direction. If $%
K_{2xy}=0$ , $K_{2z}=1$ and $g=0$ and a field $B_x$, the corresponding
magnetization curves are identical to the solid lines with $\langle
S^x\rangle $ and $\langle S^z\rangle $ exchanged.}
\end{center}
\end{figure}

If there is only a single-ion anisotropy in addition to the Heisenberg
exchange energy term, a spontaneous magnetization of the monolayer is also
induced. The upward triangles in Fig.1 show the scaled Curie temperature $%
T_c $ as a function of $S$ when there is an in-plane anisotropy in one
direction. The Curie temperatures are lower and decrease more rapidly than
those for the 3D case. The Curie points for the 2D case are lower than those
for the 3D case for the same $S$, since the number of nearest neighbours of
each site in a 2D square lattice is four, while that in a cubic 3D lattice
is six.

With an external field in addition to the single-ion anisotropies, the
magnetization will rotate. The rotation in the monolayer case exhibits
(except for the temperature scale) the same features as those of the 3D
system. If we plot the $\theta _0$, $\varphi _0-T$ curves, the qualitative
behaviour will be the same as in Fig.2. With the same parameters, the
initial angles are the same as in Fig.2. If, for example, ${\bf B}=(0.1$,
0.1, 0.1), $K_{2z}=0$ and $K_{2x}=1$, the angles ($\theta _0$, $\varphi _0$)
at zero temperature are (87.26, 3.61) degrees for $K_{2y}=0.5$ and (87.26,
2.64) degrees for $K_{2y}=0$. Above the Curie point, ($\theta _0$, $\varphi
_0$) again approach the field direction (57.74, 45) degrees, i.e. the [111]
direction.

Including both the single-ion anisotropy and DI, we plot in Fig. 1 the
scaled Curie temperature $T_c$ as a function of $S$ for the uniaxial cases $%
K_{2x}=1$ (in-plane) and $K_{2z}=1$(out-of-plane) using $g=0.018$. It is
seen that, if the easy-axis is along the out-of-plane direction, the effect
of DI is to lower $T_c$. If the easy-axis is in the in-plane direction, the
effect of DI is to raise $T_c$. This is clear from Eqs.(55) where the effect
of DI can be understood as a renormalization of the non-dispersive exchange
interaction terms:
\begin{eqnarray}
J_0\langle S^{x,y}\rangle &\rightarrow &(J_0+gT_0)\langle S^{x,y}\rangle
\text{,}  \nonumber \\
J_0\langle S^z\rangle &\rightarrow &(J_0-2gT_0)\langle S^z\rangle \text{.}
\label{57}
\end{eqnarray}
$J_{{\bf k}}$ in Eq.(55) is kept unchanged. As a result, the exchange energy
strength along the $z$-direction is suppressed and that in the $xy$-plane is
enhanced. The figure shows that in the out-of-plane case, the Curie
temperature $T_c$ decreases (compared to $g=0$) more strongly than it
increases for the in-plane case.

Now we investigate the in-plane (or easy-plane) anisotropy case, where $%
K_{2xy}=K_{2x}=K_{2y}=1$. Figure 6 shows the effective anisotropy
coefficient, defined by Eq.(51), as a function of the temperature for $S=1$,
2 and 10, when $B_z=0.1$ is taken. At zero temperature, $K_{2xy}(0)$ is the
same as the parameter $K_{2xy}$. The effective anisotropy decreases with
increasing temperature, and at the Curie point it approaches zero. In the
intermediate temperature range it is the smaller the larger the spin quantum
number $S$ is. Figure 6 also shows the equilibrium polar angle $\theta _0$
defined by Eq.(46) when $B_z=0.1$. All the properties in Fig.6 are similar
to those in Fig.3. The only difference is that the Curie temperatures in the
2D case are by more than a factor of two smaller (Fig.1) and more different
for the various spin values (Fig.6) than in the 3D case (Fig.3).

The polar angle of the magnetization depends on the temperature as well as
on the external field. It is unnecessary to plot the components of the
magnetization as functions of the external field at various temperatures as
in Fig. 4 (the 3D case) because they are qualitatively so similar.

In Refs. \cite{fro1,fro2}, the case of the out-of-plane anisotropy was
studied in detail. Since the method developed in the present paper also
enables the study of the in-plane anisotropy, it is natural to compare both
cases. We mention the following differences, which can all be explained on
the basis of Eqs. (55). The first concerns the case where there is no
external field. For an out-of-plane anisotropy, there is a magnetization
reorientation from the $z$-direction to the $xy$-plane due to DI even
without an external field \cite{fro2}. At zero temperature, the
magnetization points to the $z$-direction because of the out-of-plane
anisotropy term $K_{2z}$. With increasing temperature, the magnetization
turns to the $xy$-plane owing to the DI. In the case of in-plane anisotropy,
there is no spin reorientation in the absence of the field, because the
magnetization is already in the $xy$-plane at zero temperature.

Secondly, with the same DI strength and the same field (perpendicular to the
anisotropy direction), the magnetization reorientation temperature is
different for the two anisotropy cases as shown in Figure 7. In Fig.7(a) and
(b), the solid lines are for $S=1$ and the dotted lines are for $S=10$, both
with $g=0.018$. Figure 7(a) depicts the in-plane anisotropy case with an
external field $B_z=0.1$ along the $z$-direction, and figure 7(b) the
out-of-plane anisotropy case with an external field $B_x=0.1$ along the $x$%
-direction. The magnetization reorientation temperature of the latter case
is lower than that of the former case. This can be understood from Eqs.
(55), from which one sees that the dipole interaction enhances the effect of
the in-plane anisotropy term, whereas the opposite is true for the
out-of-plane case; i.e. the role of DI can be considered as a
renormalization of the anisotropy terms in the following sense
\begin{eqnarray}
K_{2x,y}\Phi _{x,y} &\rightarrow &K_{2x,y}\Phi _{x,y}+gT_0\text{,}  \nonumber
\\
K_{2z}\Phi _z &\rightarrow &K_{2z}\Phi _z-2gT_0\text{.}  \label{58}
\end{eqnarray}

This also explains, thirdly, the effect of the variation in the strength of
the DI, as shown in Figure 7(a) and (b). In the case of in-plane anisotropy,
increasing the DI strength hinders the reorientation from in-plane to
out-of-plane, see Fig.7(a), whereas the reorientation from out-of-plane to
in-plane is enforced in the case of the out-of-plane anisotropy. It is seen
that the DI suppresses the out-of-plane anisotropy and enhances the in-plane
anisotropy.

The fourth difference involves the variation of the external field. If an
external field $B_x$ is applied along the $x$-direction in the case of an
out-of-plane anisotropy, the magnetization reorients easier with DI than
without DI. This is shown by comparison of the dashed and solid lines in
Fig.8. The reason is that we can also interpret the effect of the DI as a
renormalization of the external field, as has been done in Ref.\cite{fro2},
see Eq. (\ref{55})
\begin{eqnarray}
B_{x,y} &\rightarrow &B_{x,y}+gT_0\langle S^{x,y}\rangle \text{,}  \nonumber
\\
B_z &\rightarrow &B_z-2gT_0\langle S^z\rangle \text{.}  \label{59}
\end{eqnarray}
This means that the effect of the DI is to enhance a magnetic field in the $%
xy$-plane, and to weaken an external field along the $z$-direction.
Therefore, if an external field $B_z$ is applied along the $z$-direction in
the case of an in-plane anisotropy, it is more difficult for the
magnetization to reorient into the out-of-plane direction with DI than
without DI. Comparison of the dotted and solid lines in Fig.8 provides an
example.

\section{Conclusions and discussion}

In this paper, we have treated a Heisenberg Hamiltonian with single-ion
anisotropies in all three directions of space by applying many-body Green's
function theory, thus generalizing previous work dealing with uniaxial
out-of-plane anisotropy. To do so, we selected the spin operators $S^x$, $%
S^y $ and $S^z$ as basis operators to construct the Green's functions
instead of the commonly used operators $S^{+}$, $S^{-}$ and $S^z$ . We
stress that in the derivation of the formalism none of the three axes is
special, so that we are always able to select a reference direction along
which the magnetization component is not zero. The key is the generalization
of the Anderson-Callen decoupling to the terms from the single-ion
anisotropy in all directions of space, Eqs.(11, 12). A formal advantage of
the use of the present operators is that the matrix governing the equations
of motion is Hermitian, whereas the ususal formalism leads to real
non-Hermitian matrices, which are more difficult to treat. A main result of
the paper is that analytical expressions can be derived for the three
components of the magnetization and the correlations $\langle (S^x)^2\rangle
$, $\langle (S^y)^2\rangle $ and $\langle (S^z)^2\rangle $ for any spin
quantum number. The formalism obtained in this paper enables a treatment of
both the in-plane and out-of-plane anisotropies.

Numerical calculations were carried out for various parameters for both 3D
and 2D systems.

For the 2D system, we included the dipole-dipole interaction. Its effect on
the magnetization can be explained on the basis of Eq.(\ref{55}), and can be
interpreted as a renormalization of the nondispersive part of the exchange
interaction terms, or of the single-ion anisotropy terms or of the external
field, see Eqs.(\ref{57}, \ref{58}, \ref{59}). The reorientation of the
magnetization from in-plane to out-of-plane under an external field occurs
more easily without the DI than with it, or in other words the DI supports
the reorientation from the out-of plane to the in-plane direction.

Because of Eq.(55), $H_z$ may appear negative. In this case the regularity
condition Eq.(28) would be meaningless because the $z$-component
magnetization should be within $0<\langle S^z\rangle <S$. Only when the
value of $g$ is under a certain value can there is a non-zero $\langle
S^z\rangle $ without a field in $z$-direction. From Eq.(58), the condition
should be $gT_0<S(K_{2z}\Phi _z-2gT_0)$. Then one should have $%
2SK_{2z}\left( 1-\frac 1{2S^2}[S(S+1)-\langle S^zS^z\rangle ]\right)
>gT_0(1+2S)$. Since $\langle S^zS^z\rangle \geq S(S+1)/3$, one has $%
g<2K_{2z}(2S-1)/3T_0(1+2S)=0.1476(2S-1)/(1+2S)$, where $K_{2z}=1$ is taken.
The upper limit increases with spin quantum number. However, because we use
scaled parameters in this paper, Eq.(52), we have $%
g<S(S+1)2K_{2z}(2S-1)/3T_0(1+2S)S(S-1/2)=(S+1)4/3T_0(1+2S)=0.2952(S+1)/(2S+1)
$. The upper limit decreases with spin quantum number. When the value of $g$
is over this limit, the $z$-component of magnetization has to be zero.
Figure 5 provides an example where the magnetization lies in the plane.

Finally, we point out that the numerical results are all calculated with
scaled parameters in order to obtain universal results, in accordance with
Refs.\cite{fro1,fro2}. With increasing $S$, the scaled exchange $J$ and
anisotropy $K_2$ become smaller as can be seen in Eq.(\ref{52}). This
results in the conclusions in Section 3 that, with increasing $S$, the
following quantities decrease: the scaled $T_c$ in Fig.1, the effective
anisotropy coefficients in Figs.3, 6 and 7, and the field under which the
reorientation occurs in Fig.4. If the parameters are not scaled but kept
unchanged, the conclusions are the other way round. A larger $S$ means a
larger magnetization. It will, therefore, result in a higher Curie
temperature. Consequently, the effective anisotropy coefficients are larger
for larger $S$ at a given temperature. For instance, at zero temperature,
the effective anisotropy coefficient should be $K_2S(S-1/2)$. A larger
magnetization is more difficult to turn by an external field. As a result,
the magnitude of the field at which the magnetization reorientation occurs
should rise with increasing $S$ for a given temperature.

$^a$ e-mail: wanghuaiyu@mail.tsinghua.edu.cn


\begin{references}
\bibitem{tyab}  S. V. Tyablikov, {\it Methods in the Quantum Theory of
Magnetism}, (Plenum, New York, 1967).

\bibitem{tahir1}  R. A. Tahir-Kheli and D. Ter Haar, Phys. Rev. {\bf 127,}
88 (1962); $ibid$. {\bf 127,} 95 (1962).

\bibitem{callen}  H. B. Callen, Phys. Rev., {\bf 130,} 890 (1963).

\bibitem{why2}  Huai-Yu Wang, Ke-Qiu Chen, En-Ge Wang, Phys. Rev. {\bf B66},
092405 (2002).

\bibitem{why3}  Huai-Yu Wang, Shan-Ying Wang, Chong-Yu Wang, Wen-Hui Duan,
and Ke-Qiu Chen, J. Phys.: Condens. Matter, {\bf 15,} 2783 (2003).

\bibitem{why4}  Huai-Yu Wang, Yun-Song Zhou, Chong-Yu Wang, D. L. Lin,
Chinese Physics, Vol.{\bf 11}, No. 2 167 (2002).

\bibitem{guo}  W. Guo, L. P. Shi and D. L. Lin, Phys. Rev. {\bf B62,} 14259
(2000).

\bibitem{moran}  T. J. Moran, J. Nogues, D. Lederman and Ivan K. Schuller,
Appl. Phys. Lett. {\bf 72,} 617 (1998); R. Jungblut, R.Coehoom, M. T.
Johnson, J. aan de Stegge and A. Reinders, J. Apply. Phys. {\bf 75,} 6659
(1994); Y. Ijiri, J. A. Borchers, R. W. Erwin, S.-H. Lee, P. J. van der Zaag
and R. M. Wolf, Phys. Rev. Lett. {\bf 80,} 608 (1998).

\bibitem{Ko97}  N.C. Koon, Phys. Rev. Lett. {\bf 78}, 4865 (1997).

\bibitem{farle}  M. Farle, W.Platow, A.N. Anisimov, P. Poulopoulos, and K.
Baberschke, Phys. Rev. {\bf B56} 5100 (1997); W. L. O'Brien, T. Droubay and
B. P. Tonner, {\bf B54} 9297 (1996); D.\ P. Pappas, K. -P. Kamper and H.
Hopster, Phys. Rev. Lett. {\bf 64,} 3179 (1990); D.\ P. Pappas, C. R.
Brundle and H. Hopster, Phys. Rev. {\bf B45} 8169 (1992).

\bibitem{usadel}  A. Moschel and K. D. Usadel, Phys. Rev. {\bf B49} 12868
(1994); A. Moschel and K. D. Usadel, Phys. Rev. {\bf B51} 16111 (1995); A.
Hucht and D. K. Usadel, Phys. Rev. {\bf B55} 12309 (1997); A. Hucht and K.
D. Usadel, J. Magn. Magn. Mater. {\bf 203} 88 (1999).

\bibitem{fro1}  P. Fr\"{o}brich, P. J. Jensen, and P. J. Kuntz, Eur. Phys.
J. B {\bf 13,} 477 (2000).

\bibitem{fro2}  P. Fr\"{o}brich, P. J. Jensen, P. J. Kuntz and A. Ecker,
Eur. Phys. J. B {\bf 18,} 579 (2000).

\bibitem{ac}  F.B. Anderson and H. B. Callen, Phys. Rev., {\bf 136,} A1068
(1964).

\bibitem{fro3}  P. Fr\"{o}brich, P. J. Kuntz, and M. Saber, Ann. Phys.
(Leipzig) {\bf 11}, No.5, 387 (2002).

\bibitem{hen02}  P. Henelius, P. Fr\"{o}brich, P.J. Kuntz, P.J. Jensen, C.
Timm, Phys. Rev. B {\bf 66}, 094407 (2002).

\bibitem{why1}  Huai-Yu Wang, Ke-Qiu Chen, En-Ge Wang, Int. J. Mod. Phys. B,
Vol. 16, No.25, 3803 (2002).

\bibitem{fro4}  P. Fr\"{o}brich, P.J. Kuntz, Eur. Phys. J. B {\bf 32}, 445
(2003).

\bibitem{Je03}  P.J. Jensen, S. Knappmann, W. Wulfhekel, H.P. Oepen, Phys.
Rev. B {\bf 67}, 184417 (2003).

\bibitem{fro5}  P. Fr\"{o}brich, P.J. Kuntz, phys. stat. sol. (b) {\bf 241},
925 (2004).

\bibitem{bog}  N. N. Bogolyubov and S. V. Tyablikov, Sov. Phys. Dokl. {\bf 4,%
} 589 (1959).

\bibitem{fro6}  P. Fr\"{o}brich, P.J. Kuntz, Phys. Rev. B {\bf 68}, 014410
(2003).

\bibitem{Millev}  Y.Millev, M. F\"{a}hnle, J. Mag. Mag. Mat. {\bf 135}, 284
(1994).

\bibitem{MW66}  N.M. Mermin, H. Wagner, Phys. Rev. Lett. {\bf 17}, 1133
(1966).
%1
\end{references}
\end{document}